\documentclass{article}
\usepackage{spconf,amsmath,graphicx}
\usepackage{array}
\usepackage{amsmath}
\usepackage{multirow}
\usepackage{setspace}
\usepackage[skip=1pt, font=normalsize]{caption}
\date{}

\usepackage{enumitem}

\usepackage{url}

\usepackage{hyperref}

\newcommand*{\Scale}[2][4]{\scalebox{#1}{$#2$}} 

\title{Scale Selective Extended Local Binary Pattern for Texture Classification}
\name{Yuting Hu, Zhiling Long, and Ghassan AlRegib}
\address{Multimedia \& Sensors Lab (MSL)\\
Center for Signal and Information Processing (CSIP)\\
School of Electrical and Computer Engineering\\
Georgia Institute of Technology, Atlanta, GA 30332-0250, USA\\
\{huyuting, zhiling.long, alregib\}@gatech.edu}
\begin{document}

\onecolumn 

\begin{description}[labelindent=1cm,leftmargin=4cm,style=multiline]

\item[\textbf{Citation}]{Y. Hu, Z. Long, and G. AlRegib, ``Scale Selective Extended Local Binary Pattern for Texture Classification'', IEEE International Conference on Acoustics, Speech and Signal Processing (ICASSP2017), pp. 1413-1417, 2017.}

\item[\textbf{DOI}]{\url{https://doi.org/10.1109/ICASSP.2017.7952389}}

\item[\textbf{Review}]{Date of publication: 5-9 March 2017}

\item[\textbf{Codes}]{\url{https://ghassanalregibdotcom.files.wordpress.com/2016/10/yuting_icassp2017_code.zip}}

\item[\textbf{Bib}] {@inproceedings{hu2017scale,\\
  title={Scale selective extended local binary pattern for texture classification},\\
  author={Hu, Yuting and Long, Zhiling and AlRegib, Ghassan},\\
  booktitle={IEEE International Conference on Acoustics, Speech and Signal Processing (ICASSP)},\\
  pages={1413--1417},\\
  year={2017},\\
  organization={IEEE}
}
}


\item[\textbf{Copyright}]{\textcopyright 2018 IEEE. Personal use of this material is permitted. Permission from IEEE must be obtained for all other uses, in any current or future media, including reprinting/republishing this material for advertising or promotional purposes,
creating new collective works, for resale or redistribution to servers or lists, or reuse of any copyrighted component
of this work in other works. }

\item[\textbf{Contact}]{\href{mailto:huyuting@gatech.edu}{huyuting@gatech.edu}  OR \href{mailto:zhiling.long@ece.gatech.edu}{zhiling.long@ece.gatech.edu} OR \href{mailto:alregib@gatech.edu}{alregib@gatech.edu}\\
    \url{http://ghassanalregib.com/} \\ }
\end{description}

\thispagestyle{empty}
\newpage
\clearpage
\setcounter{page}{1}

\twocolumn

%
\maketitle
\begin{abstract}
In this paper, we propose a new texture descriptor, scale selective extended local binary pattern (SSELBP), to characterize texture images with scale variations. We first utilize multi-scale extended local binary patterns (ELBP) with rotation-invariant and uniform mappings to capture robust local micro- and macro-features. Then, we build a scale space using Gaussian filters and calculate the histogram of multi-scale ELBPs for the image at each scale. Finally, we select the maximum values from the corresponding bins of multi-scale ELBP histograms at different scales as scale-invariant features. A comprehensive evaluation on public texture databases (KTH-TIPS and UMD) shows that the proposed SSELBP has high accuracy comparable to state-of-the-art texture descriptors on gray-scale-, rotation-, and scale-invariant texture classification but uses only one-third of the feature dimension.

\end{abstract}

\section{Introduction}
\label{sec:intro}

Texture classification as a key issue in image processing and computer vision has a variety of applications \cite{pietikainen2015two} such as content-based image retrieval, object recognition, scene understanding, and biomedical image analysis. The two main parts of texture classification are feature extraction and classification. Since feature extraction plays a relatively more important role than classifiers, researchers have done lots of work on building robust and compact texture features or descriptors.

Robustness and compactness are two conflicting goals, and a good texture descriptor should have a proper balance between them. Since texture images are commonly captured under different photometric and geometric transformations, robustness needs to consider gray-scale-, rotation-, and scale-invariances. In contrast, compactness means the descriptor should be low dimensional. One of the most popular texture descriptors is local binary pattern (LBP) proposed by Ojala et al.~\cite{ojala2002multiresolution}, which is simple but efficient for gray-scale- and rotation-invariant texture classification. Although LBP was originally proposed for texture analysis, it has been successfully used in other applications, such as face recognition and image retrieval. To improve the robustness and distinctiveness of LBP, in recent years, a large number of LBP variants have been proposed including completed local binary pattern (CLBP)~\cite{guo2010completed}, extended local binary pattern (ELBP)~\cite{liu2012extended}, and completed local derivative pattern (CLDP)~\cite{hu2016completed}. However, these LBP variants, which are robust to gray-scale and rotation variations suffer from scale variations. The work of \cite{li2012scale} introduced a feature extraction method that implements scale-invariance by estimating local scales and normalizing local regions, but has high computational complexity. To improve efficiency, Guo et al.~\cite{guo2016robust} proposed the scale-selective local binary pattern (SSLBP) that first extracts scale-sensitive local features and then applies a global operator to achieve scale-invariance. In~\cite{guo2016robust}, the scale-invariant feature extraction scheme achieves good texture classification performance on texture databases with scale variations such as KTH-TIPS~\cite{hayman2004significance} and UMD~\cite{xu2009viewpoint}. However, SSLBP as a high-dimensional descriptor has a length of $2400$.

\begin{figure*}[t]
\centering
\includegraphics[width=7in]{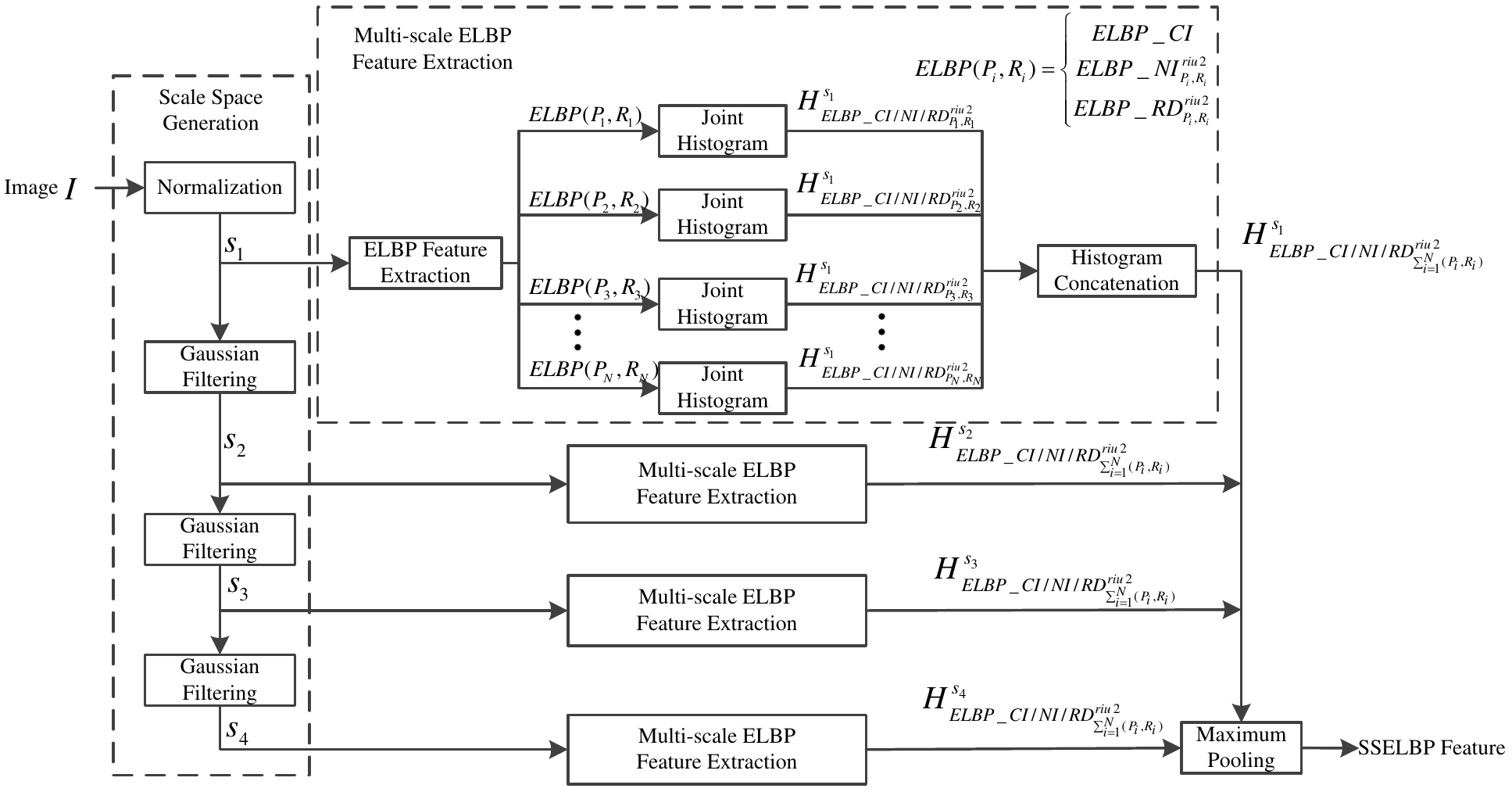}
\caption{The block diagram of the proposed SSELBP.}
\label{fig:diagram}
\end{figure*}

To reduce the feature dimension, Liu et al.~\cite{liu2012extended} proposed ELBP, which can well represent texture images and achieve good texture classification performance using limited features. Since ELBP keeps the settings of feature extraction unchanged for all images, scale variations between images may degenerate the classification performance of ELBP. To increase the robustness and efficiency of texture classification with scale variations, we propose the scale-selective extended local binary pattern (SSELBP) in this paper. We first design a framework that extracts multi-scale ELBPs. Then we build a scale space using Gaussian filters. For the image at each scale, we calculate its corresponding multi-scale ELBPs. Furthermore, we apply a maximum pooling strategy on the features across all scales and obtain scale-invariant SSELBP features. To verify the performance of SSELBP features on two texture databases with scale variations, KTH-TIPS~\cite{hayman2004significance} and UMD~\cite{xu2009viewpoint}, we use the simplest nearest neighborhood classifier (NNC)~\cite{ojala2002multiresolution} with the chi-square distance. The experimental results show that, compared to state-of-the-art descriptors, SSELBP can achieve comparable accuracy with much lower-dimensional features.

The rest of the paper is organized as follows. Section~\ref{sec:sselbp} \mbox{introduces} the proposed texture descriptor SSELBP. Section~\ref{sec:experimental results} presents experimental results on the KTH-TIPS and UMD databases. Section~\ref{sec:conclusions} makes a conclusion.

\section{The Proposed Method}
\label{sec:sselbp}

The diagram of the proposed method is shown in Fig.~\ref{fig:diagram}. \mbox{Before} we explain the blocks of this diagram in detail, we need to first briefly introduce the concept behind ELBP~\cite{liu2012extended}.

\subsection{Brief Review of ELBP}
\begin{figure}[t]
\centering
\includegraphics[width=2.5in]{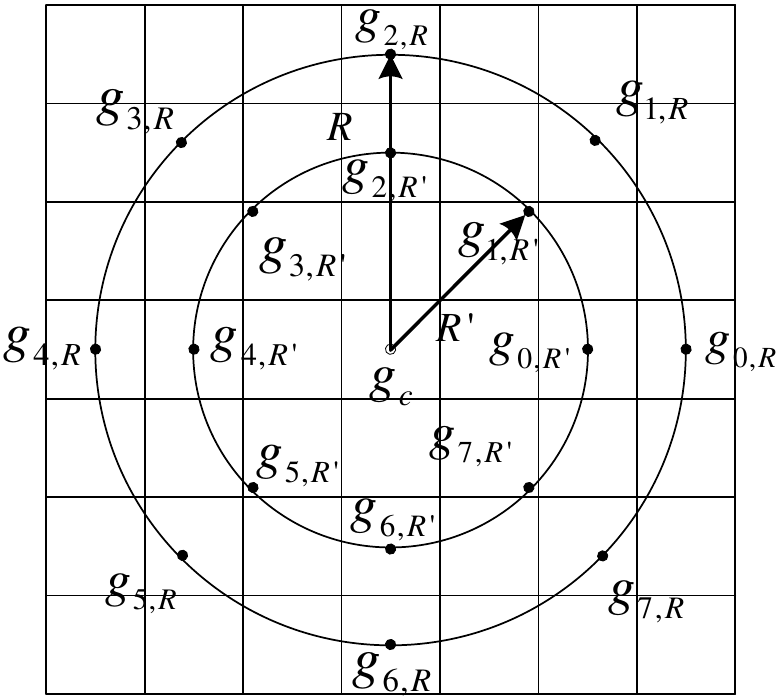}
\caption{Pixels on two circles with radii $R$ and $R'$.}
\label{fig:2circle}
\end{figure}

ELBP consists of three types of information: the intensity value of a central pixel, the intensities of pixels on a circle centered at the central pixel with radius $R$, and the intensity differences of pixels on two circles that are both centered at the central pixel and have the radii of $R$ and $R'$, $R'<R$, respectively. Given central pixel $x_c$ with intensity $g_c$, to encode the intensity information of $x_c$, operator $ELBP\_CI$ compares $g_c$ with the mean of the whole image, denoted $c_I$, as follows:
\begin{equation}
ELBP\_CI(x_c)=s(g_c-c_I),
s(x)=\left\{
    \begin{aligned}
        &1,\mbox{ if }x\geq 0\\
        &0,\mbox{ if }x<0
    \end{aligned}
    \right.
    .
\end{equation}
For each pixel in an image, $ELBP\_CI$ generates a one-bit binary pattern. In addition to $ELBP\_CI$, ELBP involves operator $ELBP\_NI$ to extract information from the intensities of neighboring pixels. In the framework of ELBP, the $P$ neighbors of a central pixel are evenly distributed on a circle with radius $R$ and have intensities denoted as $g_{p,R}$, $p=0,1,\cdots,P-1$. By comparing neighboring pixels with their average value, denoted $u_R$, $ELBP\_NI$ encodes the intensity information as follows:
\begin{equation}
\label{equ:ELBP_NI}
\begin{aligned}
&ELBP\_NI_{P,R}(x_c)=\sum_{p=0}^{P-1}s(g_{p,R}-u_R)\cdot 2^p\\
&=\sum_{p=0}^{P-1}s\left(g_{p,R}-\frac{1}{P}\sum\limits_{p=0}^{P-1}g_{p,R}\right)\cdot 2^p.
\end{aligned}
\end{equation}
Since each comparison generates one bit, $ELBP\_NI$ generates a $P$-bit binary pattern, which have been converted to a decimal value in Eq.~(\ref{equ:ELBP_NI}). The third operator involved in ELBP is $ELBP\_RD$, which encodes the intensity differences of pixels on two circles with radii $R$ and $R'$ along the radial direction. Fig.~\ref{fig:2circle} illustrates the spatial relationships of pixels on two circles. Similar to $ELBP\_NI$, $ELBP\_RD$ generates a $P$-bit binary pattern as well, and the corresponding decimal value is calculated as follows:
\begin{equation}
ELBP\_RD_{P,R}(x_c)=\sum_{p=0}^{P-1}s(g_{p,R}-g_{p,R'})\cdot 2^p.
\end{equation}
As we discussed above, operators $ELBP\_NI$ and $ELBP\_RD$ can produce $2^p$ different binary patterns. To remove the rotation effect and reduce the pattern dimension, we commonly apply rotation-invariant and uniform mappings following $ELBP\_NI$ and $ELBP\_RD$. The updated operators are denoted as $ELBP\_NI_{P,R}^{riu2}$ and $ELBP\_RD_{P,R}^{riu2}$, where superscripts ``$ri$'' and ``$u2$'' represent rotation-invariant and uniform mappings, respectively.  For example, if $P=8$, with the rotation-invariant mapping, the pattern dimension obtained from $ELBP\_NI$ or $ELBP\_RD$ can be first reduced from $2^8=256$ to $36$. Then, with the uniform mapping, the pattern dimension can be further reduced to ten. For simplification, we denote ELBP containing patterns $ELBP\_CI$, $ELBP\_NI_{P,R}^{riu2}$, and $ELBP\_RD_{P,R}^{riu2}$ as $ELBP(P,R)$ in all the following sections.

\subsection{Multi-scale ELBP Feature Extraction}

ELBP, which depends only on one or two local neighboring circles, is not robust to classify texture images with scale variations. To solve this problem, we use the multi-scale ELBP feature extraction method by involving various neighboring circles. As Fig.~\ref{fig:diagram} shows, because of different $(P,R)$ \mbox{choices}, we obtain a group of ELBPs, denoted $ELBP\left(P_i, R_i\right)$, $i = 1,2,\cdots,N$, where $N$ is determined based on the size and complexity of images. To combine patterns in $ELBP\left(P, R\right)$, we utilize the joint histogram that first concatenates patterns and then calculates the corresponding histogram. From another perspective, this combination scheme can be understood as the conversion from a joint multi-dimensional histogram to a 1-D histogram. By defining this operation as ``$/$'', we denote the joint histogram of $ELBP\_CI$, $ELBP\_NI_{P_i,R_i}^{riu2}$, and $ELBP\_RD_{P_i,R_i}^{riu2}$ as $H_{ELBP\_CI/NI/RD_{P_i, R_i}^{riu2}}$. Furthermore, we concatenate the joint histograms of $ELBP(P_i, R_i)$, $i=1,2,\cdots,N$, and yield the multi-scale ELBP feature, denoted $H_{ELBP\_CI/NI/RD_{\sum_{i=1}^{N}\left(P_i, R_i\right)}^{riu2}}$.

\subsection{Scale Space Generation}
To further increase the robustness of the proposed method on texture images with scale variations, inspired by the SSLBP framework in~\cite{guo2016robust}, we build the scale space using Gaussian filter. We first normalize image $I$ to ensure normalized image $\hat{I}$ has zero mean and unit variance. To build the scale space, we use Gaussian filters to smooth image $\hat{I}$ as follows:
\begin{equation}
s_l=
\left\{
    \begin{aligned}
    & \hat{I}, \ \quad\quad\quad\quad l=1,\\
    & s_{l-1}\ast g(\sigma), 1<l\leq L,
    \end{aligned}
\right.
\end{equation}
where $g(\sigma)$ defines a 2-D Gaussian filter with standard deviaion $\sigma$ and $L$ represents the size of the scale space. $s_l$, $l=1,2,\cdots,L$, is the image at scale $l$. With the increase of $l$, more texture details are removed and the macro-structure of the texture becomes more significant. In Fig.~\ref{fig:diagram}, we set $L = 4$ empirically for illustration.

\subsection{Maximum Pooling}
To obtain scale-invariant texture features, we adapt the idea of the maximum pooling strategy. For each scale $s_l$, we use the same $(P,R)$ set to calculate the corresponding multi-scale ELBP histogram feature, denoted $H_{ELBP\_CI/NI/RD^{riu2}_{\sum_{i=1}^{N}\left(P_i, R_i\right)}}^{s_l}$, $l = 1,2,\cdots,L$. When combining these multi-scale features, we need to consider the robustness of the pooling result on texture images with scale variations. Because of the multiple choices of $(P,R)$ in the multi-scale ELBP, we assume that the significant features of images at different scales can be captured by one parameter pair in the $(P,R)$ set. Moreover, when the scales of images change, these significant features still exist and can be captured by another parameter pair. Therefore, we utilize a maximum pooling strategy that selects the maximum values from the corresponding bins of multi-scale ELBP histogram features at different scales. The mathematically expression is shown as follows:
\begin{equation}
\label{equ:pooling}
\begin{aligned}
H&_{ELBP\_CI/RD/NI_{\sum_{i=1}^{N}\left(P_i, R_i\right)}}\\
&=\underset{l=1,2,\cdots,L}{\max}\left(H_{ELBP\_CI/RD/NI^{riu2}_{\sum_{i=1}^{N}\left(P_i, R_i\right)}}^{s_l}\right).
\end{aligned}
\end{equation}

\section{Experimental Results and Discussions}
\label{sec:experimental results}

\subsection{Texture Databases}
To test the performance of the proposed SSELBP on texture classification, we focus on two public texture databases, KTH-TIPS~\cite{hayman2004significance} and UMD~\cite{xu2009viewpoint}. The KTH-TIPS database provides totally ten texture classes. In each class, 81 samples with pose and illumination variations and varied \mbox{resolutions} ($\approx196\times 201)$ are evenly distributed in nine different \mbox{scales}. In contrast, the UMD database provides 25 classes of high resolution $(1280\times960)$ images. In each class, forty samples are captured under illumination, arbitrary rotation, significant scale, and viewpoint changes. More details about these two databases can be found in~\cite{hayman2004significance} and~\cite{xu2009viewpoint}. In these two databases, we follow the training and testing scheme in~\cite{guo2016robust} that first randomly selects half of samples in each class for training and uses the remaining half for testing. To ensure fair comparison, we repeat the training and testing process for one hundred times and calculate the average accuracy as the classification result.

\vspace{-0.15in}
\subsection{Classifier}
Since this paper mainly focuses on feature extraction rather than classifiers, we use the simplest and parameter-free classifier, the nearest neighbor classifier (NNC)~\cite{ojala2002multiresolution}, to distinguish extracted histogram features. The NNC compares a test \mbox{image} with all training images and labels the test image using the class that the training image with the highest similarity belongs to. To measure the similarity of two SSELBP histograms $T$ and $M$, which are extracted from test image $I_T$ and training image $I_M$, respectively, we use the chi-square distance as follows:
\begin{equation}
\label{eq:classifier}
\Scale[1]{
    \begin{aligned}
    D(T,M)=\sum_{w=1}^W\frac{(T_w-M_w)^2}{T_w+M_w}
    \end{aligned},
    }
\end{equation}
where $W$ is the number of bins and $T_w$ and $M_w$ are the values of $T$ and $M$ at the $w$-th bin, respectively.

\subsection{Experimental Results}
In the proposed method, to build the scale space, we use Gaussian filters with scale parameter $\sigma=2^{0.25}$ and empirically set the size of the scale space to be four. For all scales, we extract multi-scale ELBP features using the same set $(P_i,R_i)$, $i=1,2,\cdots,N$. In this paper, to reduce the feature dimension, we set $P=8$ for all radii and select $N$ radii from set $\left\{1,2,\cdots,8\right\}$. In addition, the selection of $R'$ in the calculation of $ELBP\_RD$ also depends on $R$. For example, if we use four radii $(R_1,R_2,R_3,R_4)$ to calculate the multi-scale ELBP features, the corresponding $R'$ should be $(R_0, R_1, R_2, R_3)$, where $R_0$ refers to the central pixel. To investigate the influence of $N$
on classification accuracy, we test the proposed method on the KTH-TIPS database and list all results in Table~\ref{tab:classification}. We notice that the best classification accuracy of the proposed method on the KTH-TIPS database is $98.11\%$ with radius selection $\left(2,3,4,7\right)$. When $N$ equals four or five, we can obtain higher accuracy with smaller deviations. We do not show classification accuracy when $N$ is greater than five since the average classification saturates with $N=5$. In Table~\ref{tab:classification}, the feature dimension for one radius is $2\times(P+2)\times(P+2)=200$. The increase of $N$ sacrifices the feature dimension but can not improve the classification accuracy.

We compare the classification accuracy of the proposed method with those of the state-of-the-art texture descriptors using the same classifier. Table~\ref{tab:classification2} lists the classification results and indicates the classifier each method uses. Because of the efficiency and robustness of SSLBP, we choose it as an important benchmark and compare its classification accuracy with that of the proposed SSELBP. For the KTH-TIPS database, SSELBP achieves the best accuracy $98.11\%$ among all sampling schemes, which is $0.31\%$ higher than SSLBP. For the UMD database, the classification accuracy of SSELBP is $98.96\%$, which is $0.12\%$ higher than that of SSLBP. In addition to SSLBP, we compare SSELBP with other texture descriptors such as CLBP, random projection-based feature (RP), and median robust ELBP (MRELBP).  The performance of the proposed method with a feature dimension of $800$ is comparable to state-of-the-art texture descriptors. In contrast, the feature dimensions of CLBP and SSLBP are $2200$ and $2400$, respectively.

\begin{table}[t]
\begin{center}
\caption{Classification accuracy ($\%$) of the proposed \mbox{SSELBP} using different sampling schemes on the KTH-TIPS database.}
\resizebox{0.48\textwidth}{!}{
\begin{tabular}{|c|c|c|c|c|c|}
\hline
Number of & Maximum  & Radius Selection  & Mean & Standard
&Feature\\
Radius, $N$ & Accuracy ($\%$) & for Maximum & Accuracy ($\%$) & Derivation & Dimension\\
\hline
$1$ & $96.44$ & $(2)$ & $94.80$ & $1.56$ & $200$\\
\hline
$2$ & $97.86$ & $(1,6)$ & $97.04$ & $0.63$ & $400$\\
\hline
$3$ & $98.09$ & $(2,5,8)$ & $97.51$ & $0.43$ & $600$\\
\hline
$4$ & $\mathbf{98.11}$ & $(2,3,4,7)$ & $97.71$ & $0.30$ & $800$\\
\hline
$5$ & $98.10$ & $(1,2,3,4,8)$ & $97.84$ & $0.20$ & $1000$\\
\hline
\end{tabular}
}
\label{tab:classification}
\end{center}
\end{table}

\begin{table}[t]
\begin{center}
\caption{Classification accuracy ($\%$) of the proposed \mbox{SSELBP} and state-of-the-art texture descriptors on the KTH-TIPS and UMD databases. Accuracy is originally reported. The number in the bracket following databases denotes the number of training samples used per class.}
\resizebox{0.35\textwidth}{!}{
\begin{tabular}{|c|c|c|c|}

\hline
Classification Accuracy ($\%$) & \multirow{1}{*}{\centering KTH-TIPS (40)}  &\multirow{1}{*}{\centering UMD (20)}
\\
\hline
CLBP (NNC)~\cite{guo2010completed} & $97.19$ & $98.00$\\
\hline
RP (NNC)~\cite{liu2012texture} & $97.71$ & $99.13$\\
\hline
MRELBP (NNC)~\cite{liu2015median} & - & $98.66$\\
\hline
SSLBP (NNC)~\cite{guo2016robust} & $97.8$0 & $98.84$\\
\hline
\hline
SSELBP (NNC) (Proposed) & $98.11$ & $98.96$\\
\hline
\end{tabular}
}
\label{tab:classification2}
\end{center}
\end{table}

\vspace{-0.2in}
\section{Conclusion}
\label{sec:conclusions}
We proposed a new scale-invariant texture descriptor, \mbox{SSELBP}, on the basis of ELBP. SSELBP acquired multi-scale ELBP histograms from images at all scales in the scale space generated by the Gaussian filter. SSELBP applied the maximum pooling strategy to \mbox{select} the highest co-occurrence frequency of patterns across different scales and obtain scale-invariant histogram features. In comparison with the state-of-the-art descriptors, SSELBP has at least comparable classification accuracy with much lower-dimensional features. In our future work, we will work on the automatic generation of the scale space and the automatic sampling scheme selection.

\bibliographystyle{IEEEbib}
\bibliography{refs}

\begin{thebibliography}{10}

\bibitem{pietikainen2015two}
M.~Pietik{\"a}inen and G.~Zhao,
\newblock ``Two decades of local binary patterns: A survey,''
\newblock {\em Advances in \mbox{Independent} Component Analysis and Learning
  Machines}, pp. 175--210, 2015.

\bibitem{ojala2002multiresolution}
T.~Ojala, M.~Pietik{\"a}inen, and T.~M{\"a}enp{\"a}{\"a},
\newblock ``Multi-resolution gray-scale and rotation invariant texture
  classification with local binary patterns,''
\newblock {\em Pattern Analysis and Machine Intelligence, IEEE Transactions
  on}, vol. 24, no. 7, pp. 971--987, 2002.

\bibitem{guo2010completed}
Z.~Guo, L.~Zhang, and D.~Zhang,
\newblock ``A completed modeling of local binary pattern operator for texture
  classification,''
\newblock {\em Image Processing, IEEE Transactions on}, vol. 19, no. 6, pp.
  1657--1663, 2010.

\bibitem{liu2012extended}
L.~Liu, L.~Zhao, Y.~Long, G.~Kuang, and P.~Fieguth,
\newblock ``Extended local binary patterns for texture classification,''
\newblock {\em Image and Vision Computing}, vol. 30, no. 2, pp. 86--99, 2012.

\bibitem{hu2016completed}
Y.~Hu, Z.~Long, and G.~AlRegib,
\newblock ``Completed local derivative pattern for rotation invariant texture
  classification,''
\newblock in {\em IEEE International Conference on Image Processing (ICIP)},
  2016, pp. 3548--3552.

\bibitem{li2012scale}
Z.~Li, G.~Liu, Y.~Yang, and J.~You,
\newblock ``Scale-and rotation-invariant local binary pattern using
  scale-adaptive texton and subuniform-based circular shift,''
\newblock {\em Image Processing, IEEE Transactions on}, vol. 21, no. 4, pp.
  2130--2140, 2012.

\bibitem{guo2016robust}
Z.~Guo, X.~Wang, J.~Zhou, and J.~You,
\newblock ``Robust texture image representation by scale selective local binary
  patterns,''
\newblock {\em Image Processing, IEEE Transactions on}, vol. 25, no. 2, pp.
  687--699, 2016.

\bibitem{hayman2004significance}
E.~Hayman, B.~Caputo, M.~Fritz, and J.~Eklundh,
\newblock ``On the significance of real-world conditions for material
  classification,''
\newblock in {\em European Conference on Computer Vision (ECCV)}, 2004, pp.
  253--266.

\bibitem{xu2009viewpoint}
Y.~Xu, H.~Ji, and C.~Ferm{\"u}ller,
\newblock ``Viewpoint invariant texture description using fractal analysis,''
\newblock {\em International Journal of Computer Vision}, vol. 83, no. 1, pp.
  85--100, 2009.

\bibitem{liu2012texture}
L.~Liu and P.~W. Fieguth,
\newblock ``Texture classification from random features,''
\newblock {\em Pattern Analysis and Machine Intelligence, IEEE Transactions
  on}, vol. 34, no. 3, pp. 574--586, 2012.

\bibitem{liu2015median}
L.~Liu, P.~Fieguth, M.~Pietikainen, and S.~Lao,
\newblock ``Median robust extended local binary pattern for texture
  classification,''
\newblock in {\em IEEE International Conference on Image Processing (ICIP)},
  2015, pp. 2319--2323.

\end{thebibliography}

\end{document}